\documentclass[preprint,aps]{revtex4}

\usepackage{graphicx}
\usepackage{epstopdf}
\def\depth{depth}
\def\Depth{Depth}

\def\ex{{\cal E}}
\newcommand{\PP}{{\bf P}}
\newcommand{\NC}{{\bf NC}}
\newcommand{\NP}{{\bf NP}}

\begin{document}

\title{Natural Complexity, Computational Complexity and Depth}

\author{J. Machta}
\email[]{machta@physics.umass.edu}
\affiliation{Department of Physics,
University of Massachusetts,
Amherst, MA 01003-3720, USA}
\affiliation{Santa Fe Institute, 1399 Hyde Park Rd, Santa Fe, NM 87501, USA}
\begin{abstract}
{\em \Depth}\ is a complexity measure for natural systems of the kind studied in statistical physics and is defined in terms of computational complexity.   \Depth\  quantifies the length of the shortest parallel computation required to construct a typical system state or history starting from simple initial conditions. The properties of  \depth\ are discussed and it is compared to other complexity measures.  \Depth\ can only be large for systems with embedded computation. 
\end{abstract} 
\maketitle
\section*{Lead Paragraph}
{\bf   Is there a measure of complexity that  is widely applicable and can distinguish moderately complex from very complex systems?  Parallel depth, the topic of this paper, is a candidate for such a measure. It is formulated in the probabilistic setting of statistical physics and is defined as the length of the parallel computation needed to construct a typical state of a system.   The connection between parallel depth and intuitive notions of complexity is predicated on the assumption that no system can become complex without carrying out, explicitly or implicitly, a long computation. Parallel depth is a complexity measure related to computational complexity and is essentially different from complexity measures related to entropy and information.}

\section{Introduction}
An explorer from a parallel universe arrives in the Solar System.  She makes some preliminary measurements and observes that the mass, entropy and entropy production of the system are  concentrated in one large, hot body at its center. The other much smaller objects in the system pale by comparison.   She reports back to home base that she will focus her research on the large central body and that her study will soon be completed of what appears to be a typical system on this scale and at this time epoch in this universe. 

``Have you done a complexity measurement?'' asks the team leader.

``No, I didn't think we would find anything interesting.''

``Let's do it.''

After some painstaking observations, in-depth theoretical analysis and extensive simulations, she excitedly reports her findings: the estimated complexity of this system is extremely high and it is all concentrated on the surface of one of the small, dense objects orbiting the central object. 

What did she measure?  Is there a  quantity that distinguishes the complexity of the biosphere from everything else in the Solar System?  Could such a quantity be universally defined rather than formulated using concepts from specific disciplines such as biology? The purpose of this paper is to describe one such quantity, illustrate some of its features, contrast it to some other complexity measures and argue that it does indeed formalize some of the salient intuitions about complexity.  This quantity is called `parallel \depth,'~\cite{Mac06} or simply `\depth' when there is no ambiguity.

 \Depth\ is related to computational complexity.  It is quite general in the sense that it  exists for any systems described within the probabilistic framework of statistical physics.  \Depth\ is the parallel time complexity of sampling system states or histories using an efficient, randomized parallel algorithm.  It is a formal measure of the irreducible amount of history required to create  system states from simple initial conditions and can only become large for systems with embedded computation. 
 
The notion of parallel \depth\ presented here was motivated by and is closely related to {\em logical depth} introduced by Bennett~\cite{Benn87,Benn88,Benn90,Benn95}. Both parallel \depth\ and logical depth measure the number of computational steps that are needed to generate a state.  The key difference is that the unit of measurement for parallel \depth\ is a {\em parallel} computational step whereas logical depth is defined in terms of Turing time, which is a {\em sequential} computational step.  It turns out that parallel and sequential time emphasize rather different computational resources and have rather different properties when applied as measures of natural complexity.  Although parallel \depth\ is closely related to logical depth, it is not closely related to thermodynamic depth~\cite{LlPa88,CrSh99}, which is an entropy-based measure rather than a computation-based measure.

`Complexity' has many scientific meanings.    Some of these meanings apply to specialized domains while others are rather general.   For a categorized list of proposed definitions see \cite{Lloyd01}.  Among the general definitions many are related to entropy and information.  Examples include excess entropy, effective complexity and thermodynamic depth~\cite{LlPa88,CrSh99}.   

Many systems in statistical physics have been studied from the perspective of \depth~\cite{Mac06}.   These systems include pattern formation models~\cite{Mac93a,MaGr,MaGr96}, network growth models~\cite{MaMa05}, dynamical systems~\cite{MoMa,Mac02} and spin models~\cite{MaGr96,Moore97,MoNo97}.  These studies reveal that long range correlations and power law distributions, often considered to be markers of complex systems, do not necessarily imply substantial \depth.  On the other hand, some systems in statistical physics such as diffusion limited aggregation do appear to have considerable \depth.  Diffusion limited aggregation, discussed below in Sec.\ \ref{sec:comp}, has both  long range correlations and embedded computation in the sense that its dynamics is sufficiently rich that it can be used to evaluate arbitrary  logical expressions~\cite{MaGr96}. 

 In Sec.\ \ref{sec:frame} we define \depth\ and discuss its general properties.  Section \ref{sec:exen} contrasts it to several other complexity measures via some examples.  The paper concludes in Sec.\  \ref{sec:conclusions} with a discussion.

\section{General Framework}
\label{sec:frame}
\subsection{Statistical Physics}
 \Depth\ is defined within the general framework of statistical physics and the theory of computation.  In statistical physics systems are described in terms of  probability distributions over their states.   States are broadly defined here.  A state may be a snapshot of the system at a single instant in time or it may be a sequence of snapshots constituting a time history.  States are descriptions of a system, not the system itself, and must exist within some theoretical framework.  For example, in classical mechanics, states are defined by the positions and momenta of the particles while in ecology, states might consist of numbers of individuals of each species.  

As these examples suggest, coarse-graining is often involved in specifying states.  In chemistry,  nuclei can often be described as point-like particles ignoring the hierarchy of internal degrees of freedom within the nucleus--the nucleons, quarks and gluons, strings...  In simple physical systems, a separation of time scales and energy scales often provides a rigorous justification for coarse graining.   We can ignore the nuclear degrees of freedom because the energy required to excite them is much higher than molecular energy scales in chemistry so these degrees of freedom remain in their ground states.    For more complex systems it is an art to choose an appropriately small set of degrees of freedom that adequately capture the observed behavior of a system.  The self-organization of complex systems often suggests a natural coarse-graining. 

A probabilistic description is central to our formulation.  The origin of randomness differs in different situations.  In many cases, it is enough to point to the loss of information in the coarse-graining.  The degrees of freedom that have been ignored have some influence on the behavior of the system that causes it to deviate from strictly deterministic behavior. If the coarse-graining is done well, the effects of the ignored degrees of freedom can be treated as random perturbations.  For open systems, noise from outside the system may be a source of randomness.  In any case, \depth\ is defined for probability distributions over the states of a system.

\Depth\ is defined for whole systems that are isolated or interacting simply with their environment.  For example, the Earth system is an example of a complex system that interacts with the rest of the Universe in a simple way, primarily via the exchange of blackbody radiation.  \Depth\ is not straightforwardly defined for strongly interacting components of a system even though, intuitively, these might be more or less complex.  For example, consider an agent-based model of an economic system with many complex interacting agents exchanging goods and money. It would make sense to measure the \depth\ of the entire economic system but it is not clear how to assign \depth\ meaningfully to the individual agents.

\subsection{Computation}
\label{sec:comp}
In addition to statistical physics, the formulation of \depth\ requires the theory of (classical digital) computation.     Thus, the states of the system must be expressed using a finite number of bits and the probability distributions over these states should be samplable to good approximation in finite time using a digital computer.  Our essential task is to find a computational procedure, i.e.\ a randomized algorithm, that samples the states of the system with the correct probability.  By sampling, we mean that on each run of the algorithm a description of the system is produced with the correct probability.  For example, Markov chain Monte Carlo methods such as the Metropolis-Hastings algorithm are designed to sample specific probability distributions.     The randomized algorithm makes use of random numbers and we assume that the computational device has free access to a supply of random numbers.  Finally, we assume that we have available a massively parallel computer, as described in more detail below. We minimize the parallel time to sample the states on this idealized parallel computer and define \depth\ as this parallel time.   

{\em \Depth\ is  defined as running time of the most efficient parallel algorithm for sampling the states of the system.}  Alternatively, it is the minimum number of parallel logical steps needed to transform random numbers into typical system states. 

Describing a system in terms of the computation that simulates its states allows us to make connections to other fundamental quantities related to many proposed complexity measures.  To measure \depth\ we minimize parallel time. Suppose instead one minimizes the amount of randomness needed  to sample the distribution of states.  The average number of random bits required to sample system states, using the program that minimizes this number, is roughly the entropy of the system.  The intuition for this relationship is that the Shannon entropy of an ensemble is well-approximated by the average algorithmic randomness of the states in the ensemble, see for example~\cite{Zurek89}.  It is important to note that randomness and running time cannot typically be simultaneously optimized;  a program with a short running time may not use randomness efficiently and vice versa.  Finally, if one minimizes the size of the program needed to convert  random bits into typical system states, the program size is related to the complexity measure called {\em effective complexity}~\cite{Gell95,GeLl96,AyMuSz10}.   

Since \depth\ is defined in terms of time on a parallel computer we need to briefly survey the theory of parallel computation~\cite{GiRy}.  The strategies for parallel computing can be quite different than for the more familiar sequential computing.  A simple example is adding a list of $n$ numbers.  On a single processor machine with random access memory the obvious method is to use a `for loop.'  A partial sum is initialized to zero and on each iteration  of the loop a new summand is retrieved from memory and added to the current partial sum.  The answer is produced after O$(n)$ steps, assuming each addition requires O$(1)$ steps.  An idealized machine with a single processor that can carry out simple operations such as addition and multiplication, and read and write to any cell in an arbitrarily large memory all in a single time step is known as a random access machine or RAM.  

The parallel programming method to add numbers and carry out many other computations is `divide and conquer.'  The $n$ numbers are divided into $n/2$ pairs and $n/2$ processors simultaneously add these pairs yielding $n/2$ partial sums on the first parallel step.  On subsequent parallel steps the remaining partial sums are pairwise added.  Since the number of partial sums decreases by half after each parallel step, the answer is produced in O$(\log n)$ parallel time using $n$ processors.  The same amount of computational work is needed in the parallel computation but it is distributed in a way that achieves an exponential speedup in running time relative to the single processor calculation.  

The idealized parallel machine based on the RAM is the parallel random access machine or PRAM.  The PRAM has many processors equivalent to the single processor of the RAM.  Each processor can carry out a simple computation and read and write to a global random access memory in a single time step.  The processors in a PRAM run synchronously and all processors run the same program but may branch through the program differently and read and write to different memory cells depending on their ID's.   Conflicts may arise if two processors simultaneously attempt to read from or write to the same memory location on the same step.  The way in which these conflicts are resolved yields more and less powerful variants of the PRAM.   

Because of the assumption of unit time communication with the global memory, neither the RAM nor PRAM  can be scaled up indefinitely since they must ultimately violate either speed of light constraints for fixed component size or materials constraints on miniaturization.  
Nonetheless, the PRAM  is useful for understanding strategies for parallel algorithms  that can be adapted for more realistic machines.  Since our purpose is to understand to what extent the simulation of the system can be broken into independent tasks we can ignore communication time.

The theory of parallel computational complexity is concerned with the limits to the efficiency with which problems can be solved in parallel on a PRAM.  For an introduction to the field see, for example~\cite{GiRy}.  It is formulated as a scaling theory and the basic question that is asked is how do computational resources scale with the problem size.  The primary resources are parallel time and hardware.  Interesting results follow if the parallel time is minimized under the constraint that  number of processors is restricted to grow as a power of the problem size. Under this reasonable constraint, which is the assumption that is adopted here, some problems, such as addition, are efficiently parallelizable and others are apparently not.  Addition, for example, requires a number of processors that grows linearly in the problem and can then be solved in logarithmic time. 

Computational complexity theory defines broad classes of problems.  The first such complexity class is \PP; the class of problems that can be solved on a PRAM in polynomial time with polynomially many processors.  Addition of $n$ numbers is in \PP\ but it is also within the subclass \NC\ of problems that can be solved in polylogarithmic time, i.e.\ a power of the logarithm of the problem size, using polynomially many processors.  
Problems in \NC\ are considered to be efficiently parallelizable.  While it is obvious that $\NC \subseteq \PP$, it is not clear whether the inclusion is strict.  The question is whether there exist problems in \PP\ that cannot be solved in polylogarithmic time in parallel.  Such problems would be `inherently sequential.' This question has not been settled but the consensus is that there exist problems in \PP\ that are, in fact, inherently sequential.  Candidates for such problems are the set of \PP-complete problems.   The reader is referred to~\cite{GrHoRu} for a discussion of the theory of \PP-completeness and a compendium of \PP-complete problems.  The definition of \PP-completeness and the question of whether $\NC = \PP$ closely parallels the definition of \NP-completeness and the question of whether $\PP = \NP$.  It is strongly believed that both inclusions $\NC \subseteq \PP$ and $\PP \subseteq \NP$ are, in fact, strict.  Suffice it to say that no polylog time parallel algorithm has been found for any \PP-complete problem. 

The class of \PP-complete is defined via `reductions'~\cite{GrHoRu}.  Roughly speaking, a problem ${\cal A}$ reduces to a problem ${\cal B}$ if an algorithm for ${\cal B}$ can be used as a subroutine to solve ${\cal A}$ with only polylog additional parallel time needed.  A problem is \PP-complete if all other problems in \PP\ can be reduced to it.  Thus, if  a polylog time parallel algorithm were to be found for one  \PP-complete problem, then efficient parallel algorithms would exist for all problems in \PP\ and the class \PP\ would collapse to \NC.  It is interesting to note, however, that there exist random ensembles of \PP-complete problems and their solutions that can be  sampled efficiently in parallel~\cite{MaDeMeMo11} illustrating the point that it is sometimes easier to simultaneously generate a problem and its solution than to first generate the problem and then solve it.

The definition of parallel time sketched in the foregoing was presented in the context of a specific model of parallel computation, the PRAM, and thus it does not appear to be a satisfactory basis for a general definition of natural complexity.  However, computational complexity theory is not tied to a specific model of computation.  The broad complexity classes \NC, \PP, and \NP\  can be equivalently defined in terms of PRAMs, Turing machines or Boolean circuits.  For Boolean circuits, the resource that is equivalent to parallel time is circuit depth, motivating the choice of the term `\depth' for the complexity measure discussed here.  Circuit depth is the length of the longest path from inputs to outputs in a Boolean circuit.  Finally, the theory of descriptive complexity~\cite{Im99} shows that computational complexity theory  can be formulated in terms of the logical description of computational problems without reference to computing machines at all.   The quantity corresponding to parallel time in a first-order logical description of a problem is the number of quantifier alternations.  The robustness of computational complexity theory makes it a suitable candidate for a general and fundamental definition of physical complexity.

\begin{figure}[t]
\includegraphics[width=0.45\textwidth]{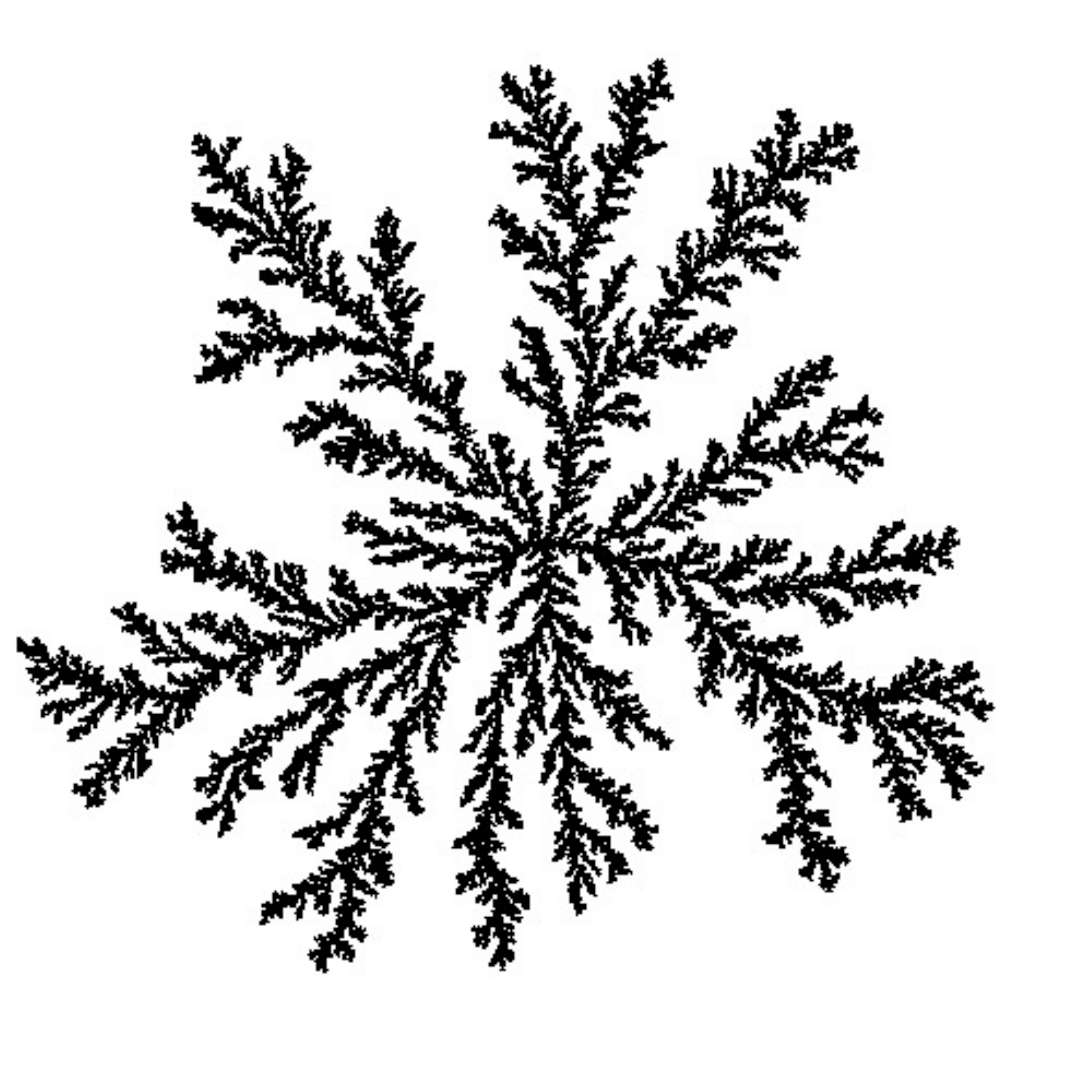}
\caption{A typical diffusion limited aggregate.  This aggregate has $N=1\times 10^6$ particles.}
\label{fig:dla}
\end{figure}

Parallel computational complexity theory and the complexity classes defined above refer to the computational difficulty of solving problems whereas \depth\ is defined in terms of the difficulty of sampling distributions.  We can illustrate this distinction in the context of an important model in statistical physics, diffusion limited aggregation (DLA).  The DLA model defines a probability distribution on patterns in two (or more) dimensions according to the following growth law.   Initially the aggregate consists of a single  particle fixed at the origin of space.  A second  particle is initialized some distance from the origin and performs a random walk.  If  during its random motion it touches the initial  particle it sticks to it and the aggregate now consists of two particles.  If it walks too far from the origin, it is removed from the system and a new particle started.  The aggregate grows when a diffusing particle touches the existing aggregate and sticks to the particle it touches.  The aggregate grows, adding one particle at a time, until it consists of $N$ particles.  A typical aggregate with $N=1\times10^6$ particles is shown in Fig.\ \ref{fig:dla}.  As can be seen in the figure, DLA produces visually complex, fractal structures.   

It is important to observe that the DLA growth rule is  sequential in the sense that  only  one walking particle is in the system at a time.  If many particles walk at the same time the ensemble of aggregates is very different. Thus one might guess that DLA is inherently sequential in the computational sense that it will require O($N$) parallel steps to create an aggregate of $N$ particles and that the \depth\ of DLA is O($N$).   We can think of the process of creating the aggregate as converting  random numbers first into random walk trajectories and then, using the growth rules, into the aggregate itself.  Thus the growth rules define a sampling algorithm for DLA.  We can also think of this algorithm as solving a problem.  The problem is converting an input, the random numbers controlling the random walk trajectories, into an output, the aggregate.  We can ask how hard that problem is to solve in parallel.  It turns out that our intuition about the sequential nature of the process is confirmed in the sense that we can show that the  DLA problem is \PP-complete~\cite{MaGr96}.  Thus there is almost certainly no polylog time parallel algorithm for solving the problem of converting random numbers into random walks and then into aggregates of the type shown in Fig.\ \ref{fig:dla}.  

The \PP-completeness result is nice and strongly suggests that DLA has considerable \depth.  Unfortunately, there is a catch.  \Depth\ is defined as the complexity of sampling DLA  using the {\em most efficient} parallel algorithm.  The defining growth law is only one way to create DLA aggregates.  Other methods are known and one of these has also been shown to be associated with a \PP-complete problem~\cite{Mac93a}. However, to prove that DLA has more than polylog \depth\ we would need to consider {\em all} possible sampling problems for DLA.  It would be nice to have a well-developed theory of the complexity of sampling but unfortunately this does not yet exist.  Thus, we do not have tools available to set rigorous lower bounds on \depth.  On the other hand, it is straightforward to set an upper bound on \depth\ by actually demonstrating a parallel sampling algorithm and determining its running time. The defining growth law for DLA shows that its \depth\ is no worse than linear in $N$.  In \cite{TiMa04} we  improve this result slightly.

Because it is defined in terms of parallel computation,  \depth\ is  {\em maximal}~\cite{Mac06}.  Maximality is a generalization of the property of intensivity in statistical mechanics.  Properties such as temperature, pressure and chemical potential in statistical mechanics are {\em intensive} because they are  independent of  system size for homogeneous equilibrium systems.  Such systems  can be decomposed, to good approximation, into independent subsystems each of which has the same value of the intensive quantities.    Now consider a system that is not homogeneous but can still be decomposed into  independent subsystems.  If the subsystems have different \depth s then the  \depth\ of the whole is the maximum of the \depth s of the subsystems.  This property follows trivially from the independence of the subsystems  and the fact that \depth\ is running time on a parallel computer.  We can assign a separate set of processors to each subsystem and the \depth\ of the whole system is then determined by when the last set of processors has finished.   Note that if the system is homogeneous so that the subsystems are independent and identically distributed then \depth\ is also intensive. 

\section{\Depth\ and Other Complexity Measures}
\label{sec:exen}
Two interesting and well-studied  complexity measures are  excess entropy and statistical complexity \cite{CrPa83,Grass86, CrYo89,BiNeTi01,BiNeTi01a}.   The domain to which these measures most naturally apply is to stationary time series, although generalizations to other domains are possible.  For a stationary time series we may define $H_T$ as the entropy of a block of the series of length $T$,
\begin{equation}
\label{ }
H_T = - {\bf E}_P  \log P_T(S)
\end{equation} 
where $P_T(S)$ is the probability of time series  $S=(s_1,\ldots,s_T)$ of length $T$ where each element of the series is chosen from a set of symbols  and  ${\bf E}_P$ is the expectation with respect to $P$.
The entropy rate $h$ is defined as
\begin{equation}
\label{ }
h=  \lim_{T\rightarrow\infty} \frac{H_T}{T} .
\end{equation}
This limit must converge from above and the sum of the deviations from the limit is the excess entropy.  Specifically, let $h_T=H_{T+1}-H_T$ and let $H_0=0$.  Then the excess entropy $\ex$ is given by
\begin{equation}
\label{ }
\ex=\sum_{T=0}^\infty (h_T-h)
\end{equation}

The statistical complexity~\cite{CrYo89} requires knowing the causal states of the system as defined in computational mechanics.  Causal states are sets of past states with statistically indistinguishable futures.  The statistical complexity is the entropy of the collection of causal states.  It can be shown that the excess entropy is a lower bound for the statistical complexity. 

A time series that has no memory, e.g.\ a series of independent coin tosses, has no excess entropy and no statistical complexity.  Let $s_i$ be $1$ with probability $1/2$ and $-1$ with probability $1/2$.  The block entropies are given by $H_T = T H_1= T \log (2)$ for all $T \geq 0$ as required by the additive property of entropy so $h=h_T= \log 2$ and the excess entropy vanishes.  Since the past is uncorrelated with the future, there is only one causal state consisting of all possible pasts so the statistical complexity also vanishes.    Since each $s_i$ is independent, one processor can be assigned to each $s_i$. This processor simply reads a random bit and assigns it to $s_i$.  The entire series is thus created in a O(1) parallel steps independent of the length of the series. The depth of this random series is O(1).  

Next consider a coin toss with memory.  The set of symbols is again $\{-1,1\}$.  Let $s_{n+1}=s_n$ with probability $p$ and $s_{n+1}=-s_n$ with probability $1-p$.  It is straightforward to compute the block entropies to be $H_T= \log 2-(T-1)\left[p \log p + (1-p) \log (1-p)\right]$ and from this expression we find that the excess entropy is
\begin{equation}
\label{ }
\ex= \log 2 +\left[ p \log p + (1-p) \log (1-p)\right].
\end{equation} 
Although $\ex$ varies with $p$, the statistical complexity is independent of $p$ and given by $\log 2$.  The two equally likely causal states are distinguished by the value of $s_0$ since the statistical properties of the future, $s_1, s_2, \ldots $, is completely determined by $s_0$.  As advertised, the statistical complexity is here greater than the excess entropy.

To estimate the \depth\ of the coin toss with memory we seek an efficient parallel algorithm for simulating it. Here is one of several possible methods. First, in parallel construct a biased sequence of bits  $u_1, u_2, \ldots u_{T}$ where each $u_i$ is independently chosen to be $+1$ with probability $p$ and $-1$ with probability $1-p$.    The $u_j$ implement the flipping of successive coins with probability $p$ according to relation $s_j=u_j s_{j-1}$. Since the $u_j$ are independent, they can be constructed in parallel time O(1).  Next choose $s_0$ randomly.  Finally, let $s_i= s_0 \prod_{j=1}^i u_j$.  As in the case of addition, the product can be carried out in logarithmic parallel time using divide and conquer.  Thus the \depth\ of the coin toss with memory is O($\log T$).

It is interesting to note that  the excess entropy of the coin toss with memory is computed directly from the probability distribution of blocks but estimating the \depth\ requires additional understanding of how one might simulate the time series.  We chose one method and determined the parallel time for implementing it but did not prove that it was the best method.  We have an upper bound rather than an exact answer.  This open-endedness is a feature of \depth\ that we will return to in the next section.

\Depth, excess entropy and statistical complexity are all small for the coin toss with memory even though the correlation time (average length of blocks of like bits) is $1/\log p$ and diverges as $p$ approaches one. For the coin toss with memory, there is a finite excess entropy that approaches $\log 2$ as $p$ approaches either zero or one.   Of course, for $p=0$ or 1, the time series is simple and deterministic.  For $p=1$ all bits are the same and for $p=0$ the bits alternate.  The \depth\ is a constant for these  deterministic time series.  

An interesting example of a stochastic time series with  excess entropy that diverges as a power law is described in the paper in this volume by Debowski~\cite{De11} in the context of linguistics and called the Santa Fe process.  The Santa Fe process produces a series of pairs $s_i=(K_i, z_i)$ for $i=1,\ldots, T$. The $K_i$ are independent, identically distributed  positive integers chosen from a power law density, $p(K_i=k) = A(\beta) k^{-1/\beta}$ with $0<\beta<1$ and $A(\beta)$ the appropriate normalization. The $z_i$ are not independent but are associated with $K_i$ according to $z_i=Z_{K_i}$.  For each positive integer $M$, $Z_M$ is an independent coin toss ($Z_M=-1$ with probability $1/2$ and $Z_M=+1$ with probability $1/2$).  Of course, we only need to know $Z_M$ for the values of $M$ that appear in the list $K_1, \ldots, K_T$.  As $T$ increases the number of distinct values of $K$  increases as $T^\beta$.

For the Santa Fe process, the excess entropy diverges as a power of time~\cite{De11},
\begin{equation}
\label{ }
\ex \sim T^\beta
\end{equation}
The Santa Fe process has excess entropy that diverges as a power law  because, as the sequence is made longer, information encoded in the set  $\{Z_M\}$  is slowly revealed.  It is important here that the probability density for $K$ decays slowly so that the amount of memory (number of distinct $Z_M$ bits in the sequence) diverges as a power law in the length of the sequence.

Although the excess entropy and statistical complexity of the Santa Fe process both diverge as a power law, the \depth\ is logarithmic.  Here is a way to sample the Santa Fe process of length $T$ quickly in parallel.  First the sequence $\{K_i | i=1,\ldots, T\}$ is constructed.  Since the $K_i$'s are independent random variables chosen from a power law distribution, the construction can be done independently with one group of processor assigned to each $i$,  $0<i<T$.  This step requires O($\log T$) parallel time because the integers that need to be manipulated to generate the random $K_i$ grows as a power of $T$.  

Another group of processors generates a sequence of random bits $\{z_i | i=1,\ldots, T\}$. This can  be done independently for each $i$ in O(1) parallel time.  This sequence is not correct however because of possible repeated values of the $K_i$.  We  eliminate dupicates and implicitly determine the independent random bits $Z_M$ in the following two parallel steps. We use an auxiliary variable $u_i$ that is initialized to zero for all $i$, $0<i \leq T$.  For all pairs $0<i,j \leq T $ in parallel compare $K_i$ and $K_j$.  If $K_i=K_j$ then let $m=\max(i,j)$ and write $u_m \leftarrow 1$. (Note that many processor may simultaneously attempt to write to $u_m$ but they will all write 1.  We have implicitly assumed the CRCW COMMON PRAM model where many processors may simultaneously write to the same memory element if they all agree.) This step requires O(1) parallel time using $T^2$ processors. After this step is complete  $u_i=0$ if and only if either $K_i$ is distinct or $i$ is the least index among the set of duplicates of $K_i$.

One more parallel step is required to correctly determine the set $\{z_i\}$.  For all pairs $0<i,j \leq T $ in parallel if $K_i=K_j$ and if $u_i=0$ then $z_j\leftarrow z_i$  else if $u_j=0$ then  $z_i\leftarrow z_j$. This step also requires O(1) parallel time using $T^2$ processors. Although we did not explicitly construct $\{Z_M\}$, after this step is complete we have for each $M$ for which it is defined, $Z_M = z_\ell$ such that $\ell$ is the least index $i$ for which $K_i=M$.  Since this $z_\ell$ was not overwritten in the above step, it is an independent random bit, as required.  In addition, for all $j>\ell$ such that $K_j=K_\ell$ then $z_j=Z_{K_j}$ as required.

We have demonstrated a parallel algorithm whose running time is O ($T$) using $T^2$ processors.  Thus the \depth\ of the Santa Fe process is no greater than logarithmic in $T$.

The Santa Fe process is an example where the  \depth\ is much less than the excess entropy.  The Santa Fe process has lots of memory and long range correlations but almost no embedded computation.  The contrast between \depth\ and excess entropy for the Santa Fe process highlights a key difference between \depth\ and entropy-based measures of complexity.    Entropy-based measures are sensitive to stored information and how it is distributed in a system but not whether it has been processed by a long computation.  \Depth\ is sensitive to embedded computation and can only be large for systems that carry out computationally complex information processing.

Are there examples of bit sequences with \depth\ that increases as a power of the length of the sequence?  A candidate from statistical physics is the equilibrium one-dimensional Ising model with long range interactions.  The one-dimensional Ising model with nearest neighbor interactions is equivalent to the coin toss with memory.  It does not have a finite temperature phase transition and has little complexity.  On the other hand, Ising models with long range interactions have a finite temperature phase transition~\cite{Dyson69,BrMoYo86,FiHu88} and long range order.  The only known  ways to sample the states of this system are to use Markov chain Monte Carlo methods such as the Metropolis algorithm  or cluster algorithms of the Swendsen-Wang variety~\cite{SwWa}.  It is believed that these algorithms, even in parallel, require polynomial running time to converge to equilibrium at critical points.  The excess entropy of one-dimensional Ising models with long range interactions also diverges as power law in the system length $T$~\cite{BiNeTi01} though apparently for quite different reasons.

We saw in the previous section that \depth\ is maximal.   Excess entropy, by contrast, is {\em sub-extensive}.  In statistical mechanics, an extensive property is one that grows linearly in the number of degrees for freedom for homogeneous systems with short range correlations.  A sub-extensive property grows more slowly than the number of degrees of freedom. The sub-extensivity of the excess entropy  follows simply from its definition since excess entropy is that part of the entropy that remains after the entropy rate, or linearly growing part, is subtracted off.   

The maximal property of parallel \depth\ has interesting consequences for a system with one small but very complex component weakly interacting with a much larger environment.   The maximal property insures that the \depth\ of the whole system is dominated by the complex subsystem.   Extensive or sub-extensive measures of complexity both suffer from the drawback that the complex subsystem will make a small contribution to the complexity of the whole.  More formally, consider a system consisting of $R+1$ independent subsystems of which $R$ are not very complex as measured by a complexity measure while one is much more complex.  Let's assign a complexity $\mu_0$ to each of the $R$ less complex subsystems and $\mu_1$ to the single much more complex subsystem such that $\mu_0 \ll \mu_1$. Logical depth, excess entropy and many other measures  are  additive for independent subsystems so that the total complexity $\mu$ would be given by $\mu=R\mu_0 + \mu_1$.  If $R$ is sufficiently large then $\mu \approx R\mu_0$ and the complex system makes an insignificant contribution to the complexity of the whole.  This problem is not alleviated by constructing an intensive measure by dividing by $R+1$.  On the other hand, for a maximal measure  $\mu = \mu_1$, independent of $R$ and the very complex subsystem is clearly identified in its much larger environment.  

The story of the explorer from a parallel universe visiting the Solar System illustrates the advantage of using a complexity measure with the maximal property.  Suppose the explorer used an additive complexity measure instead of one that has the maximal property.  For the sake of argument, let's also assume that the conditions of the previous paragraph apply: the Sun can be well approximated as a  large number of nearly independent subsystems such that the Earth's complexity is much greater than each subsystem but much less than the sum over all the subsystems comprising the Sun.  The  complexity of the Solar System would then be dominated by the Sun and would  not be significantly  different than the complexity of a lifeless planetary system with a similar central star.  The additive complexity measure would have failed to alert the explorer to the existence of the Earth's biosphere. This argument suggest that good measures of complexity should have the maximal property and should not be additive.

Excess entropy, statistical complexity and parallel \depth\ all assign the highest complexity to systems  that lie between completely ordered and completely random.  All assign greater complexity to systems with long range correlations and are able to distinguish simple, long range correlations (e.g.\ the coin toss with memory aka the nearest-neighbor one-dimensional Ising model) from more complex correlations   (e.g.\ the long range one-dimensional Ising model).  For some examples such as the Santa Fe process, the long range correlations are complex as measured by excess entropy and statistical complexity but simple as measured by  \depth.   On the other hand, there are presumably time series with \depth\ that grows  linearly whereas the excess entropy is, by definition, sublinear.

\section{Discussion}
\label{sec:conclusions}
\Depth\ has been defined as the length of the shortest path on a parallel computer from random numbers to typical states (or histories) of a system.   \Depth, like entropy, is defined for any system described within  the general, probabilistic framework of statistical physics and both are functions of the probability distribution of system states.  \Depth\  reflects some of our intuitions about complexity. Like other reasonable measures of complexity, it is small for systems that are either very ordered or very disordered and tends to be largest at transitions between order and disorder.  Systems with considerable \depth\ are capable of computation.  Unlike other complexity measures, \depth\ is maximal, that is, it is dominated by a complex subsystem even if it is part of a much larger but less complex environment.  

Emergence and complexity are closely related concepts.  Emergence is both a signature of complexity and a means for reducing it.
The relation between emergence and complexity is apparent even in simple examples taken from elementary physics.  Newton's laws, if applied blindly,  equate \depth\ with physical time.  In the absence of any other knowledge, we  simply have to integrate Newton's laws without taking shortcuts.  The simulation will take a time that is proportional to the physical time.  As we develop theoretical understanding we can take shortcuts.  Consider the simple pendulum.  Once we discover the exact solution for the simple pendulum we can avoid explicitly integrating the equations of motion to sample typical trajectories.  Like the alternating series 1010101..., the \depth\ of simple pendulum trajectories is small.  Pendulum trajectories are not complex because they are ordered.  

Equilibrium statistical mechanics provides a second example of the interplay of emergence and complexity. Consider a classical gas confined for a long time in a closed, isolated container.  Again, in the absence of other knowledge, we would have to integrate the equations of motion for as long as the gas is in the container and \depth\ appears to grow in proportion to physical time.  However, once we discover that such a system rapidly reaches thermodynamic equilibrium, we see that its \depth\ is much less than its age.  It is not necessary to do a full integration of the many-body equations of motion. Instead,  we can do a relatively short molecular dynamics or Monte Carlo simulation to sample an equilibrium state, independent of how long the gas is in the container.  Equilibrium states (at least away from critical points) are not complex because they are disordered in the sense that they have maximum entropy given the constraints on the system.  In both of these examples, new properties emerge.  For the pendulum, periodic motion emerges and for the gas, thermodynamic equilibrium emerges.  In both cases, knowing the emergent behavior reduces the estimate of \depth\ relative to a blind application of the fundamental dynamics.  

Our estimate of \depth\ is thus contingent on our understanding of a system.
This phenomenon is dramatically illustrated by the scientific history of simulations of the Ising model.  In 1987, after more than three decades of numerical studies, a new algorithm was developed by Swendsen and Wang~\cite{SwWa} that significantly accelerates the simulation of critical states of the  Ising model~\cite{NeBa}.  The Swendsen-Wang algorithm can be efficiently parallelized and thus  provides a much better upper bound on the \depth\ of Ising critical states.  It is interesting that the improved Swendsen-Wang algorithm is able to accelerate the simulation of the Ising model because it identifies and acts on transient large scale  structures that are present at the critical point.  Older algorithms, such as the Metropolis algorithm~\cite{Met}, flip a single spin at a time and therefore do not ``know'' about the relevant critical fluctuations.  These algorithms correctly sample the critical state but they do so blindly and inefficiently.  On the one hand, large scale critical fluctuations are the signature of the complexity of the critical state.  On the other hand, understanding the dynamics of the critical fluctuations allows us to show that the \depth\ of the critical state is less than we previously believed though it is still the most complex point in the phase diagram of the Ising model.  

The Ising model again illustrates the general phenomenon of `emergence' and its connection to complexity.  Single spin flip dynamics can be viewed as  the microscopic or `fundamental' dynamics of the Ising model.  Single-spin flip dynamics is correct and very general but inefficient at critical points because here large scale structure `emerges' and has its own slow dynamics.  Identifying and understanding the dynamics of the emergent structures permits more efficient sampling using the Swendsen-Wang algorithm.

The Ising model example also reveals a serious objection with using \depth\ as a complexity measure.  It is not computable and  can only be bounded from above.  As we understand a system better we may find an algorithm that simulates the system more efficiently than was previously believed to be possible.  We would then see that the \depth\ of the system is smaller than we previously thought.
Is the non-computability of \depth\ a feature or a bug?  I conjecture that it is a necessary feature of any general measure that has a hope of distinguishing differences in complexity among very complex systems.  
Scientific progress is open-ended and presumably not computable. Measures of the complexity of the objects of scientific study should have the same property.  This observation should not be a cause for despair, many systems are already well-understood and major advances in understanding them are not likely.  For such systems we can have good estimates of \depth.

The above remarks take us full circle.  The original intent was to construct an intuitively plausible, fundamental general measure of complexity not tied to specific disciplines. In one sense this has been accomplished since \depth\ is defined in the very general  setting of statistical physics and computational complexity theory and has at least some of the properties we expect of a complexity measure.  
On the other hand, to actually estimate \depth\ requires discipline-specific knowledge. The estimate is provisional and subject to revision as understanding of the system advances. 

Unlike \depth, there is a simple expression for entropy given a probability distribution though actually computing the entropy may be  difficult.   Entropy-based complexity  measures such as excess entropy are therefore likely to be blind to many features of very complex systems.  They may detect various correlations and memory but cannot distinguish whether the information embodied in these features  arose from computationally simple or computationally complex processes. 

What are complexity measures good for?  What more do we learn by measuring complexity that we wouldn't otherwise know? This question is especially acute for parallel \depth\ and logical depth since these quantities cannot be measured without first developing considerable understanding of the system.  A complexity measure by itself is not very useful or interesting.  Instead, complexity measures should play a central role in a  theory of complex systems.  Such a theory should organize and classify  systems within its purview.  It should enhance our understanding of specific systems and help make predictions or formulate hypotheses.  It should allow us to go beyond what we might learn by studying  systems in isolation.  
The next step in developing the field and demonstrating the utility of \depth\  and other complexity measures is to make non-trivial deductive statements relating these measures to other  characteristic properties of complex systems.

\section{Acknowledgments}
This work was supported in part by NSF grant DMR-0907235.  I thank the Santa Fe Institute for supporting visits during which much of this paper was written, and for sponsoring the January 2011 workshop, ``Randomness, Structure, and Causality:
Measures of Complexity from Theory to Applications,'' which was the inspiration for this paper. I thank Cris Moore for useful discussions and suggesting improvements to the parallel algorithms.

\end{document}